# Screened non-bonded interactions in native proteins manipulate optimal paths for robust residue communication

**Ali Rana Atilgan, Deniz Turgut, Canan Atilgan***

Faculty of Engineering and Natural Sciences, Sabanci University, 34956 Istanbul, Turkey,

**\***Corresponding author: canan@sabanciuniv.edu

**ABSTRACT**

A protein structure is represented as a network of residues whereby edges are determined by intra-molecular contacts. We introduce inhomogeneity into these networks by assigning each edge a weight that is determined by amino-acid pair potentials. Two methodologies are utilized to calculate the average path lengths (APLs) between pairs: To minimize (i) the maximum weight in the *strong* APL, and (ii) the total weight in the *weak* APL. We systematically screen edges that have higher than a cutoff potential and calculate the shortest APLs in these reduced networks, while keeping chain connectivity. Therefore, perturbations introduced at a selected region of the residue network propagate to remote regions only along the non-screened edges that retain their ability to disseminate the perturbation. The shortest APLs computed from the reduced homogeneous networks with only the strongest few non-bonded pairs closely reproduce the strong APLs from the weighted networks. The rate of change in the APL in the reduced residue network as compared to its randomly connected counterpart remains constant until a lower bound. Upon further link removal, this property shows an abrupt increase, towards a random coil behavior. Under different perturbation scenarios, diverse optimal paths emerge for robust residue communication.

# INTRODUCTION

Interactions, delay, and feedback are the three key characteristics of complex fluids. Using these features, entities at different time and length scales communicate with great accuracy, efficiency and speed (1). Self-assembling molecular systems are complex fluids with robust and adaptable architectures. Proteins, whose internal motions are decisive on their folding, stability, and function, are exquisite examples of these. Proteins are under constant bombardment in their environment – e.g. in the cell where other small and large molecules are densely and heterogeneously distributed, or in the test tube with only water around, displaying ceaseless fluctuations around their folded structure. Since proteins function efficiently, accurately and rapidly in the crowded environment of the cell, they are expected to be effective information transmitters by design. The fact of the protein being functional or not depends on the size of these fluctuations and how they are instilled, making use of the concerted action of residues located at different regions of the protein (2-5). It is, therefore, of utmost interest to investigate how proteins respond to changes in the environment under physiological or extreme conditions.

The response of any structure to perturbations depends on its general architecture. For proteins, local, regular packing geometries (6) cannot provide short distances between highly separated residues for fast information transmission. In fact, it has been shown that random packing of hard spheres similar to soft condensed matter is observed in a set of representative proteins (7). Consistent with the concurrent requirement of order and randomness in the protein structure, we (8) and others (9-11), have recently shown that proteins are organized within the small-world network topology. A network is referred to as *small-world* if the average shortest path between any two vertices scales logarithmically with the total number of vertices, provided that a high local clustering is observed (12). Such properties are common in many real-world complex networks (13,14), and there are examples from a diverse pool of applications such as WWW (15), the internet (16), math co-authorship (17), power grid (12) and residue networks (8).

In recent years, we treated proteins as networks of interacting amino acid pairs to determine their network structure and to identify the adaptive mechanisms in response to perturbations (8,18,19). In fact, similar network treatments of proteins predict collective domain motions, hot spots, and conserved sites (2,20-23). For these networks we used the term "residue networks" (8) to distinguish them from "protein networks" which are used to describe systems of interacting proteins (24). We carried out a statistical analysis to show that proteins may be treated within the small-world network topology. We analyzed the local and global properties of these networks with their spatial location in the three-dimensional structure of the protein. We also showed that the shortest path lengths in the residue networks and residue fluctuations are highly correlated. In the past few years, the network treatment of residues in proteins have been adopted to study their various features such as conserved long-range interactions (25), functional residues (26,27), protein-protein association (28), and detection of structural elements (29).

In all these treatments, which have been successful in describing many important properties of proteins and provide insight as to how they function, the identities of individual amino acids are omitted in the calculations. In other words, specificity is taken into account in an indirect manner, by assuming that the locations of the different amino acid types along the contour of the polymeric chain have been operational in determining the particular average three-dimensional structure. In this viewpoint, the interactions between different pairs, triplets, etc. of amino acids are assumed to be smeared out, and the observed behavior once the protein is folded, is driven by the overall structure. In fact, it has been noted that the residue non-specific interactions contribute more to the overall stability of proteins by a factor of about five, compared to distinct residue-residue interactions (30). The question remains, however, as to the extent to which such a coarsened description of the folded protein may be used to determine other crucial properties, especially those pertaining to dynamics.

In this study, we further elaborate on the paths between residue pairs, which we term information pathways, to understand how they relate to dynamic phenomena in proteins. In particular, it is of interest to understand allosteric interactions mediated through the changes in the dynamic fluctuations around the average structure, both in the presence and absence of conformational changes, the latter having very recently been shown to exist in proteins through a series of NMR experiments (31). To this end, we attribute weights to the links between residue pairs using knowledge-based potentials (32,33), and discuss the relationship between dynamic phenomena occurring in proteins and the optimal path lengths obtained from these weighted networks. We show that it is possible to extract minimal sub-graphs from the fully connected networks of residues, where a few designed-in interactions overlaying the backbone are sufficient to display communication path lengths similar to that of the full residue network. We also demonstrate an application of these ideas using a non-redundant data set of interacting proteins, and extract residue pairs on the interface of the receptor/ligand that frequently appear along information pathways.

**METHODS**

**Spatial residue networks.** For the single protein calculations, we utilize 595 proteins with sequence homology less than 25 % (34) and sizes spanning ca. 50 to 1000 residues; this protein set is identical to that used in our previous study of residue networks (8). For the receptor-ligand complexes, on the other hand, we use the non-redundant benchmark set of Weng and collaborators developed for testing docking algorithms that contains an overall of 59 pairs of proteins with 22 enzyme-inhibitor complexes, 19 antibody-antigen complexes, 11 other complexes, and seven difficult test cases (35). We form spatial residue networks from each of these proteins using their Cartesian coordinates reported in the protein data bank (PDB) (36). In these networks, each residue is represented as a single point, centered on the $C_\beta$ atoms; the $C_\alpha$ atoms are used for Glycine residues. Given the $C_\beta$ coordinates of a protein with $N$ residues, a contact map can be formed for a selected cut-off radius, $r_c$, an upper limit for the separation between two residues in contact. This contact map also describes a network which is generated such that if two residues are in contact, then there is a connection (edge) between these two residues (nodes) (8). Thus, the elements of the so-called adjacency matrix, **A**, are given by

$$A_{ij} = \begin{cases} \mathcal{H}(r_c - r_{ij}) & i \neq j \\ 0 & i = j \end{cases} \quad (1)$$

Here, $r_{ij}$ is the distance between the $i$th and $j$th nodes, $\mathcal{H}(x)$ is the Heaviside step function given by $\mathcal{H}(x) = 1$ for $x > 0$ and $\mathcal{H}(x) = 0$ for $x \leq 0$. We adopt the value for the cutoff distance $r_c = 6.7$ Å that includes all neighbors within the first coordination shell around a central residue.

In the case of the weighted residue networks, we assign weights to the edges according to the inter-residue interaction "potentials" of Miyazawa and Jernigan (32) and Thomas and Dill (33). These are statistical potentials extracted from a protein data base. Both potentials have been extensively tested in threading algorithms (37,38), protein stability and designability studies (39), folding and binding energetics, as well as amino acid classification (40). The Miyazawa-Jernigan (MJ) potential is based on a set of protein subunit structures exceeding 1600 in number (32). In their treatment of the problem, the system is taken as an equilibrium mixture of unconnected residues and effective solvent atoms. The Bethe approximation is employed to estimate the contact energies from the numbers of contacts that arise in the sample. Excluded volume is taken into account by the inclusion of a hard-core repulsion between the residues and a repulsive packing-density-dependent term. The Thomas-Dill potential, on the other hand, utilizes a much smaller data set of 37 proteins (33). The authors use the folded chain conformation as the reference state, instead of a collection of randomly mixed particles of residues and solvent molecules [in treatments using the Bethe approximation, the problem of reference states has been addressed and corrections have been proposed (41)]. Thomas and Dill employ an iterative method which extracts pair potentials that incrementally drive the system towards a lowest energy structure that corresponds to the native structure. The main discrepancies in the statistical potentials that result from the approximate treatment or neglect of excluded volume, chain connectivity and interdependence of pairing frequencies are therefore intrinsically taken care of.

In this study, we have repeated all the calculations using both the Miyazawa-Jernigan and the Thomas-Dill knowledge-based potentials. Despite differences in details, the main results and conclusions reached do not change with the choice of potential. In what follows, we therefore report only results from the Thomas-Dill potentials. We assign $e_{ij}$, value of the connection between the $i^{th}$ and $j^{th}$ residue, according to the inter-residue interaction potential between the $i^{th}$ and $j^{th}$ residue types. Thus, the links connecting the residue pairs with the least favorable interaction energy have the lowest weight, i.e. the highest value.

**Network descriptors.** The networks are classified by local and global parameters, all of which can be derived from the adjacency matrix (eq. 1). In the absence of edge weights, the most general descriptors of the network structure are average connectivity of a node, and the average shortest path length through the network. The connectivity $k_i$ of residue $i$ is the number of neighbors of that residue, $k_i = \sum_{j=1}^{N} A_{ij}$. The average connectivity of the network is thus $K = <k_i>$, where the brackets denote the average over all nodes. The connectivity distribution of the residue networks follow the Gaussian distribution (8).

The shortest path length, $L_{ij}^{h}$, of a homogeneous network, where the links have no weights, is the average over the minimum number of connections that must be traversed to connect residue pair $i$ and $j$. In computing the shortest path between a pair of nodes, we make use of the fact that the number of different paths connecting a pair of nodes $i$ and $j$ in $n$ steps is given by $(\mathbf{A}^n)_{ij}$. Thus, the shortest path between nodes $i$ and $j$ is given by the minimum power, $m$, of $\mathbf{A}$ for which $(\mathbf{A}^m)_{ij}$ is non-zero.

In the presence of weights, it is possible to redefine the path lengths so as to take into account the skewing effects of the weights. Weights may be factored into the path lengths using different optimality criteria. We define two criteria for paths between two residues (42-44), *weak disorder* and *strong disorder*. In the former one, the optimal path connecting residues *i* and *j* is the length of the path, $L_{ij}^w$, that minimizes the sum of the weights along the path. We employ Dijkstra algorithm to compute the optimal paths in the weak disorder case. In the latter (strong disorder) case, $L_{ij}^s$ is the length of the shortest path that minimizes the maximum weight along the path. To obtain $L_{ij}^s$, we sort the links in descending order and sequentially remove the links beginning with the highest weight (lowest energy). We continue to remove the links until we find the bottleneck link which will cause the connectivity between vertices *i* and *j* to be lost. We then compute the length of this remaining path in terms of the number of intervening links. Note that once the optimal path connecting residues *i* and *j* is determined, the path length is simply the sum of the connections along the path; i.e. the step lengths themselves are not weighted.

The characteristic path length of the network is the average,

$$L^\dagger = \frac{2}{N(N-1)} \sum_{i=1}^{N-1} \sum_{j=i+1}^{N} L_{ij}^\dagger \qquad (2)$$

where the dagger symbol, †, represents the homogeneous, weak or strong paths, $L^h$, $L^w$, and $L^s$, respectively. Note that $L^\dagger$ is a measure of the global properties, reflecting the overall efficiency of the network, under the imposed constraints; i.e. the lower $L^\dagger$ is, the faster information is communicated through the network.

**RESULTS**

**Random coils as a basis for comparison.** Proteins may be modeled as networks where a special set of interactions are imposed on chain connectivity and the extent to which such interactions are specially designed are of interest here. In this study, we generate a variety of networks based on selected proteins. A firm basis for comparing the various networks that may be formed from a given chain with a known contact number is a chain of the same length and the same number of connections for each of its nodes, but a randomized set of links between the nodes. To generate such networks, we rewire every residue (node) randomly to another residue chosen from a uniform distribution such that each residue has the same number of neighbors (contact number, $k_i$), while the contact order changes; chain connectivity is preserved by keeping the (*i*, *i*+1) contacts intact. Such a network corresponds to the random coil conformation of a polymer chain at an arbitrary point in time. In our previous study, it was established that the proteins have a Poisson distribution of contacts (8). It is also known from network theory that a completely random, Poisson distributed network has the shortest path length,

$$L_{random} = \frac{\log N}{\log K}. \qquad (4)$$

As shown in figure 1 bottom curve, it is verified that the randomized chains behave exactly as expected from a completely random collection of nodes. Average path lengths on the residue networks, $L^h$, on the other hand, are significantly higher than the randomized networks while still preserving the approximately logarithmic dependence on number of residues, as shown with the filled circles in figure 1. The loss of high optimality (i.e. a two-fold increase in the shortest path lengths compared to a random network) must be compensated by the emergence of functionality in the self-organized structure. This exchange is achieved along the scaffold of the non-random networks formed by the residues of the proteins.

**Optimal paths in the presence of weights.** In the absence of weight information of the links (i.e. for a homogeneous network), $L^h$ is the only parameter we can use as a measure of the distance between nodes in the network with $N$ vertices. In the presence of weights, the heterogeneity of the medium is taken into account; hence different types of optimality criteria can be defined. In the case of *weak disorder,* the sum of the potentials along the optimal path is minimized to obtain $L^w$. This can be interpreted as the path that causes minimum possible total disturbance to the residues along the path. The links with lower potentials are more likely to tolerate the disturbances. In Figure 1 we display a comparison of shortest paths of homogeneous and weak disordered networks, $L^h$ (symbols) and $L^w$ (line), respectively, with that of the random coil. The correlation between the two data sets is excellent, showing that the weighted network in the weak disorder limit behaves similar to the homogeneous network. The optimal path in the *strong disorder*, on the other hand, is the path that minimizes the maximum of the potentials along the path, which can be interpreted as the shortest path that causes minimal maximum disturbance along the path. As exhibited in Figure 1 for the strong disorder case (see the open circles and the overlaying best-fitting dashed line), $L^s$ is significantly larger than $L^w$ by an average factor of 1.3.

**Are weights imposed on the links significant for the protein?** To answer this question, we randomly reassign the potentials attributed to pairs of residues. This is achieved by redistributing the 210 different types of pair potentials in the Thomas-Dill potential matrix, so that the same residue type pair always has the same value. As such, the underlying network structure remains unchanged, while the optimal paths that are preferred will be affected. The results based on these networks are obtained from five realizations of this randomization.

Two major observations are made for such networks: In the weak disorder limit, the optimal path lengths increase (data not shown), signifying that the residue pairs are specially distributed in the protein network so as to have similar allotment of weights around a given node, although the values themselves have a large span [-1.8 … 1.5]. Moreover, the strong paths in the weight-randomized networks are longer (shown by the dashed line in figure 1), further corroborating this finding with the more stringent constraint that key links minimizing the maximum weight along given paths exist in the folded protein.

**Identifying redundancies in the protein communication pathways by extracting sub-networks.** We deduce sub-networks from the original residue networks of each of the 595 proteins utilized in this work by systematically removing links that have values higher than a given cut-off value, $e_{cut}$. Chain connectivity is preserved regardless of the residue types flunking a given bond. We rely on the fact that, a protein under external disturbance will have a higher tendency to lose communication through high energy contacts, while the low energy ones will be more cohesive. The shortest path lengths of each of the remaining networks are subsequently computed. Several important cases are presented in figure 2, as a function of the random coil of the same size, $N$, and the same original number of neighbors, $K$ (equation 4). The distribution of the links is shown in the inset to this figure, and the chosen cut-off values are marked on the distribution.

The redundancy in the proteins is such that, when ca. half of the non-bonded contacts are disregarded, $e_{cut} = 0$, the system still has the same shortest path length as the full protein that preserves all of its contacts (compare the green line and the black data points). Upon further removal of links, the paths get longer, and they overlap with $L^s$ at $e_{cut} = -0.6\ k_BT$ (compare the blue line and the red data points). At this point, only ca. 20% of the long-range contacts remain in the sub-networks. Further removal of contacts results in a sudden increase in the shortest path lengths, exemplified by the case of $e_{cut} = -1.0\ k_BT$. In figure 1, this data set is shown in purple, along with the best fitting line (slope = 22.6, in comparison to the random networks where the slope is one). Note also that the scatter in the data is extreme, signifying that the logarithmic dependence of path lengths on number of residues is lost.

Another way to observe this data is by plotting the shortest path lengths of the sub-networks as a function of the random coil of the same size, $N$, and the modified (reduced) number of neighbors, $K'$ (figure 3). Although the path length increases as networks with less contacts are formed, as expected, the slope of the best-fitting line remains constant until $e_{cut} = -0.6\ k_BT$, i.e. coincides with the original, fully connected network that utilizes the strong paths as was shown in figure 2. Further removal of links results in a dramatic increase in the shortest paths, as exemplified by the $e_{cut} = -1.0\ k_BT$ case (purple; values on the right y-axis). Again, it is observed that the scatter in the data increases as the sub-networks approach a linear chain ($e_{cut} = -1.8\ k_BT$, i.e. only connectivity remains).

**DISCUSSION**

A folded protein needs to perform its function under the constraints that the overall shape is suitable for the task it undertakes, while it is not energetically penalized. As a molecular machine, it needs to optimize the time it takes to communicate the incoming information, which, to a first approximation, may be assumed to be linearly dependent on the shortest path length in its residue network. Excluded volume imposes another limit on the size of the molecule. As incoming information, we refer to perturbations that are imparted on one or several of the residues. Changes in the environmental conditions that are reflected on thermodynamic parameters, such as the temperature, will affect the whole system. The latter are not of concern in this study, since these may potentially change the overall network structure.

In the previous section we have displayed results that introduce several different perspectives to evaluate how folded proteins are organized so as to manage their redundancies under sub-optimal conditions. Our basis for comparison is the random coil, whereby a Poisson distributed arrangement of residues will always lead to the most optimal path length, given by the analytical relationship of equation 4. The random networks constructed for figure 1 have the same average number of neighbors as their folded network counterparts [$K = 6.9$, as shown in ref. (8)]. They may be thought of as compact chains that constantly change their partners at different points in time. They, therefore, represent an average over many significantly different configurations, in direct opposition to the case of a folded protein, where residues always keep the same neighbors while they fluctuate in space. For a given amount of excluded volume, decided upon by chain connectivity and the number of long-range contacts, the random coils give a limiting value for how fast information may be spread through the system.

On the other hand, information spreading will take on different forms in a protein depending on the type of local perturbation that is received. Two limiting situations may be distinguished: (i) Proteins experience constant random fluctuations from the environment under the usual conditions they function; e.g. random collisions with solvent molecules, formation of local hot spots, etc. We classify these perturbations, extensive in number but small in the size of fluctuation they invoke, as "everyday events." (ii) At other times, there will be large perturbations that will be targeted on specific regions, such as those occurring during binding, or approach of a large cellular body to unspecified regions of the protein. We classify these perturbations as "extreme events." The modes of response from the protein are expected to be different for the two types of events. In other biological systems, such modified reactions to different types of input (global vs. pathway specific noise) were also observed and quantified; e.g. for the variation in the behavior of genetically identical cells (45,46).

In folded proteins, the network structure, equivalent to a coarse graining obtained from the average conformation of the folded structure, is expected to remain nearly the same under both conditions. However, the way the energy will be transmitted throughout the network will differ according to the type of perturbation. Noting that the network is mostly made up of residues held together by non-bonded interactions, the proximity of pairs of residues will not differ; e.g., in many cases, the structure of the bound and unbound forms of a ligand protein to its receptor is less than the experimental uncertainty as in the case of chymotrypsin inhibitor II (2). However, the transfer of information (energy) along the residue network will only occur if the fluctuations in neighboring residues are correlated along any chosen pathway [as conformational variability increases, the communication of a signal in a molecule, e.g. conductance, occurs with less strength and over a broader range of values, as was recently demonstrated through unique experiments in a series of diphenyl containing small molecule systems (47)]. For small perturbations caused by random fluctuations, the correlations between neighboring residues are expected not to be affected, and the most probable pathway for information transmission is the lowest energy one – i.e. $L^w$. For large impacts (extreme events), although the overall network structure will be preserved due to the pressure exerted by the compact structure of the molecule, the correlations between pairs of residues that are weakly connected to each other will be lost. For the purpose of information propagation, those pathways may be assumed to be non-existent; i.e. those network connections will be lost.

**Properties of the residue network under varying degrees of external perturbations.** Usually, the impacts imparted on the protein in its usual environment will be intermediate between the two extremes of small perturbations and large impacts. Our analysis in figure 3 shows the operational limits of these molecular machines: We may classify those perturbations that delete nearly half the non-bonded contacts from being functional (i.e. $e_{cut} = 0.0\ k_BT$) as everyday events. The change in the average path length of the protein relative to the change in that of the randomly rewired counterpart ($\partial L'/\partial L'_{random}$, where $L'$ refers to path length on the sub-networks with the lower average connectivity, $K'$) remains fixed for that range (figure 3). The latter quantity is shown for the whole range of values of $e_{cut}$ in figure 4a. In the same range of values, the average shortest path length, a size dependent quantity, is also constant (figure 4b). The change in the average number of neighbors of a node is also relatively small, decreasing from 6.2 to 5 (figure 4c). Noting that two of these neighbors are located along the chain, at $e_{cut} = 0.0\ k_BT$ an average node has lost one of its four non-bonded neighbors.

Further removal of the links signifies even larger perturbations to the protein. Up to ca. $e_{cut} = -0.6$ $k_BT$, where the shortest path lengths on the sub-networks coincide with the strong paths of the original weighted residue networks (marked by the dashed lines in figures 4a-c), the quantity $\partial L'/\partial L'_{random}$ shows a decreasing trend (inset to figure 4a). In the range of $e_{cut} = -0.6 - 0.0\ k_BT$, the increase in $L$ is less than a factor of two for all sizes of proteins, whereas its value increases logarithmically beyond that cut-off ($e_{cut} < -0.7\ k_BT$; see figure 4b). The logarithmic dependence of the path length on chain size is also preserved in this range (see figures 2 and 3). Note that at this critical value of the cut-off, only about one non-bonded contact per average node remains (figure 4c).

Representative proteins of α, β, α/β types are shown in figure 5; ribbon diagrams of the structures deposited in the protein data bank are shown in the first column. All non-bonded contacts (thin lines) superimposed on the backbone (thick lines) are shown in the second column. The strongest links that form the underlying structure and that give the polymeric chain its protein-like path lengths are shown in the third column. Any other interactions added to these create redundancies that contribute to the robustness of the structure so that the protein is able to function under the harsh conditions of the cell. In reality, depending on the size and direction of the impact, some of the weaker links that are located far from that site may be preserved; i.e. we do not expect the links to be lost hierarchically. Nevertheless, the protein's reaction to the perturbation, as measured by the average path lengths of the effectively remaining contacts, is relatively insensitive to size and direction, as long as the most cohesive of the interactions remains intact.

**Practical application: Optimal paths in interacting proteins.** We postulate that residues, frequently found along the paths connecting a receptor – ligand pair, control the communication between the two proteins. Since binding is an event that requires exchange of large amounts of energy, in this treatment, we use the optimal paths with strong disorder which emphasize the largest barriers to be crossed along the way. Using the benchmark set of 59 receptor-ligand complexes (35) described in the Methods, we seek the pairs of residues that are most significant in determining key interactions. In the data set, there are ca. $2\times10^6$ such pathways, giving a statistically significant number for our analysis.

We first record the pairs that form bridges between receptor and ligand for every path that originates in the receptor and ends in the ligand; i.e. residue $i$ is located on the receptor and residue $j$ is located on the ligand and they are connected within the network formed by the protein – protein complex. We then take into account the fact that the propensity of a selected amino acid type being located along the interaction surface significantly varies, as reported by Ma et al. (48); e.g. TRP, ARG and GLN are the residues that are found most frequently on the interface. Therefore, we normalize the probability of finding a residue pair along the strong pathways, $p_{i\leftrightarrow j}$. Thus, the conditional probability, $p(i\leftrightarrow j\ |\ i, j)$, can be computed by relating the probability that the pair actually appears along the selected paths, to the probability of each of the residues in the pair being located on the interface, $q_i$ and $q_j$:

$$p(i \leftrightarrow j \mid i, j) = \frac{p_{i\leftrightarrow j}/(q_i q_j)}{\sum p_{i\leftrightarrow j}/(q_i q_j)} \qquad (5)$$

$p_{i \leftrightarrow j}$ is assumed to be proportional to the frequencies that these pairs are observed in the interface along the strong paths determined in this study. $q_i$ and $q_j$ are taken to be proportional to the propensity of the residue to be found in the interface of either the ligand or the receptor, as reported in the literature (48). The resulting conditional probabilities of the most significant pairs are listed in Table 1, along with the value of the TD contact potential.

Note that the pairs that are used in the paths consist mostly of the hydrophobic-hydrophobic interaction types, though not necessarily appearing in the order of cohesive energy. In fact, if all amino acids are grouped in the broadest sense of hydrophobic, polar, charged, and GLY, over 42% of all pairs that appear along the interface and that are on the strong paths make hydrophobic-hydrophobic contacts. Furthermore, the interactions need not be symmetric; in fact, the most significant pairs have ILE on the receptor and VAL on the ligand (normalized probability is 0.13). The reverse arrangement does not appear to be significant. A similar observation is also made for the ALA – ILE pair. In contrast, ILE and LEU pairs appear to be involved in specific interactions, though not with a significant preference for the ligand or the receptor. One example ligand-receptor system of α-chymotrypsin in complex with eglin c is shown in figure 6. Residue pairs that are on the largest number of pathways between the receptor and the ligand are shown in orange and green, respectively. Note that in the large interaction surface of the protein pairs, it is possible to identify four key interactions utilizing three residues on one protein and four on the other.

**FINAL REMARKS**

In this study, we have taken a network perspective of analyzing proteins, and have shown that residue specificity plays an important role in protein functioning. A statistical analysis on nearly 600 non-homolog proteins has led us to define key quantities for discriminating the underlying structure that make the protein robust in the environment where it is functional. In particular, the quantity $\partial L'/\partial L'_{random}$ (figures 3 and 4a) has been uniquely defined for finding a critical threshold value to determine the key interactions in the protein, if it is to survive extreme events, and to continue carrying out its function. Our results also support the finding that optimized protein sequences can tolerate relatively large random errors in pair potentials obtained using a variety of methodologies (41,49). In fact, none of our conclusions change when the work here is repeated with the pair potentials of Miyazawa and Jernigan (32), rather than that of Thomas and Dill (33), although there are differences in the details of, e.g. figure 4.

In this work, we propose that in events involving small perturbations, the total energy to traverse that path will be important, and information will flow through the optimal paths with weak disorder, similar to that in the homogeneous network. On the other hand, when large perturbations are involved, such events require surpassing the largest energy barriers along the paths. In the current approach, the same pair potentials are used as thermodynamic measures in the former case, and as kinetic measures in the latter. If a pair of residues has high contact energy, it may be assumed that the energy that must be used to separate them will be commensurate with its value, to a first approximation. Due to other effects such as the size and the shape of the residues, slight modifications may be included. So far as one realizes the network approach used here involves many approximations, as well as a large amount of coarse graining overlaying the atomic structure, such an approach has firm grounds. The strong paths, therefore, set a limit on the protein whereby the robust structure resists large amounts of external perturbations and preserves its protein-like communication pathways.

It has been suggested that an unfolded protein may fold when a sufficiently large number of contacts that are consistent with the native structure are initiated (50), and our findings are also in agreement with this viewpoint, providing one simple approach to detecting those key contacts. Furthermore, using this approach, we have been able to define key residues that form bridges between interacting proteins (Table 1). Note that nearly half the surface area of the total protein, and therefore an overwhelming number of residue pairs, is involved in protein-protein interactions. The few key contact pairs may be used as primary links in identifying the interaction geometry, overlaid by the energy lowering contributions from the rest of the pairs in solving protein-protein interaction problems.

## ACKNOWLEDGEMENTS

C.A. acknowledges support by the Turkish Academy of Sciences in the framework of the Young Scientist Award Program (CB/TÜBA-GEBIP/2004-3).

**Table 1.** Residue pairs that appear in the interface with significantly enhanced probabilities.

| Residue Pair (Receptor → Ligand) | Propensity-normalized probability, $p(i \leftrightarrow j \mid i, j)$ | Contact Potential (units of $k_BT$) |
|---|---|---|
| ILE – VAL | 0.13 | -0.98 |
| ALA – ILE | 0.041 | -0.64 |
| ILE – ILE | 0.039 | -0.71 |
| ILE – LEU | 0.036 | -1.04 |
| GLU – LYS | 0.032 | -0.09 |
| LEU- ILE | 0.030 | -1.04 |
| VAL – VAL | 0.027 | -1.15 |

**Figure Captions**

*Figure 1.* Optimal path lengths, $L^h$ (●), $L^w$ (solid line), $L^s$ (○), of the protein networks in comparison to those of the theoretical value of Poisson distributed random networks of the same size and number of neighbors ($L_{random}$, eq. 4). Results are presented for the non-redundant set of 595 proteins whereby values for proteins of size (m±1)×10; m = 3,5,… are averaged. Protein path lengths computed with the weak disorder limit are not distinguishable from those of shortest paths on homogeneous networks; both may be best-fitted by a line of slope 5.2. Optimization with the *strong* criterion results in networks with significantly longer path lengths (best-fitting line through the data is shown by the dashed line; slope is 9.0). For comparison, random coils have also been generated by random rewiring of the residue networks while preserving connectivity (see text). These networks provide the same result as a totally randomized network (no chain connectivity) of the same size (slope is 1.0). At the other extreme, randomized weights have been imposed on the original residue networks (dotted line). $L^s$ for these are longer by a factor of ca. 1.3, indicating that the weights in a protein are specifically distributed.

*Figure 2.* Optimal path lengths of the protein networks constructed with various schemes as a function of the randomized counterparts of the original networks (eq. 4). Sub-networks from the original residue networks are deduced using the edge values, whose distribution for the 210 possible residue pair interactions are shown in the inset. Edges with values higher than a given cut-off, $e_{cut}$, are removed and the new *shortest* path lengths of these sub-networks are computed; connectivity is preserved. The redundancy in the proteins is such that, when ca. half of the long-range contacts are removed, the system still has the same path length. Upon further removal of contacts, the paths get longer, and they overlap with $L^s$ at $e_{cut}$ = -0.6 $k_BT$ (only ca. 20% of the long-range contacts remaining). Further removal of contacts results in a sudden increase in the shortest path lengths, exemplified by the case of $e_{cut}$ = -1.0 $k_BT$ (slope = 22.6).

*Figure 3.* Optimal path lengths of the protein networks constructed with various schemes as a function of the randomized counterparts of the newly constructed networks, $L'_{random}$ = log $N$ / log $K'$. Subnetworks are formed as described in the caption to Fig. 2. Although the path length increases as networks with fewer contacts are formed, the slope of the best-fitting line remains constant until $e_{cut}$ = -0.6 $k_BT$, i.e. coincides with the original, fully connected network that utilizes the strong paths. Further removal of links results in a dramatic increase in the shortest paths, as exemplified by the $e_{cut}$ = -1.0 $k_BT$ case (purple; values on the right *y*-axis). Also notice that the scatter in the data increases as the subnetworks approach a linear chain ($e_{cut}$ = –1.8 $k_BT$, i.e. only connectivity remains).

*Figure 4.* Change in network parameters of the sub-networks formed as described in the caption to Fig. 2, with cut-off imposed on the link values, $e_{cut}$ so as to include the screening effect: **(a)** For a wide range of $e_{cut}$, the slopes of the curves of Fig. 3, $\partial L'/\partial L'_{random}$, remain nearly constant. Once ca. 85 % of the non-bonded contacts are removed, there is a sudden increase in the slopes. A close-up look at this range in the inset shows that there is a dip in the slopes prior to this departure from protein-like behavior. **(b)** Change in sub-network shortest path lengths with $e_{cut}$ for different protein sizes. The differences between the logarithms of the path lengths for different network sizes remain constant until the transition region of $e_{cut}$. (c) Dependence of chain connectivity on $e_{cut}$, which is commensurate with the distribution of the link values (inset to Fig. 2).

*Figure 5.* Example networks from proteins with common folds. Columns represent the ribbon structure, total networks, and "strong" networks. In the network representations, the backbone traces are shown by the thicker lines, and the non-bonded contacts are shown by the thin gray lines. 14, 21, 13, and 18 % of the non-bonded contacts remain in these proteins, pdb codes 1cgn, 1i1b, 1igd, and 1tim, respectively.

*Figure 6.* Example receptor – ligand system of the enzyme eglin c in complex with the inhibitor α-chymotrypsin; PDB code: 1acb. Bridging residue pairs that are on the largest number of pathways between the receptor and the ligand are shown in orange and green, respectively. The interacting pairs are (enzyme – inhibitor): PHE39 – TYR49, PHE41 – LEU47, VAL213 – LEU45, TRP215 – LEU45; note that LEU45 interacts with two residues.

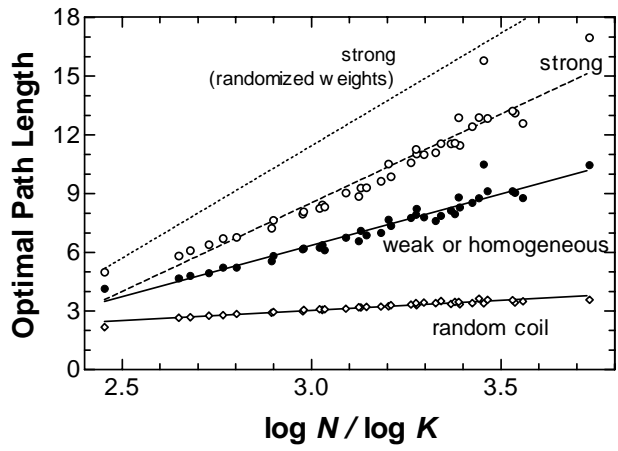

Figure 1

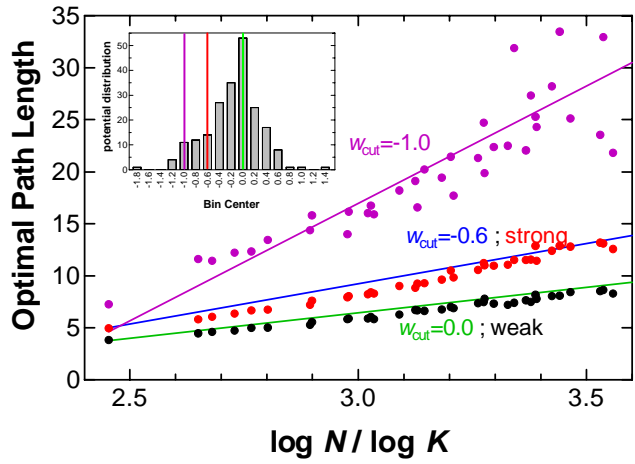

Figure 2

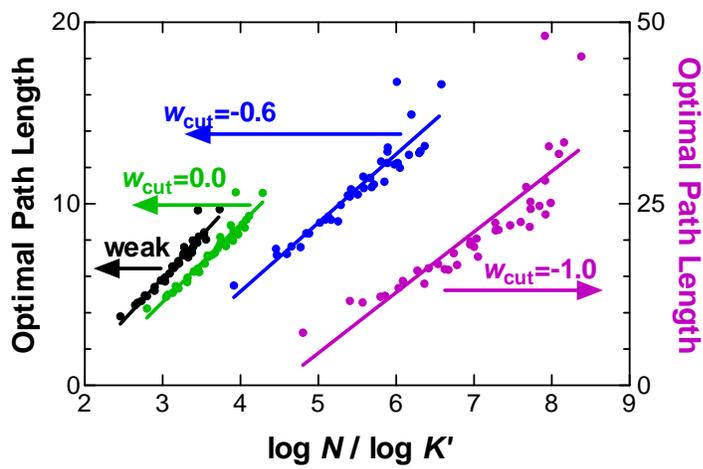

Figure 3

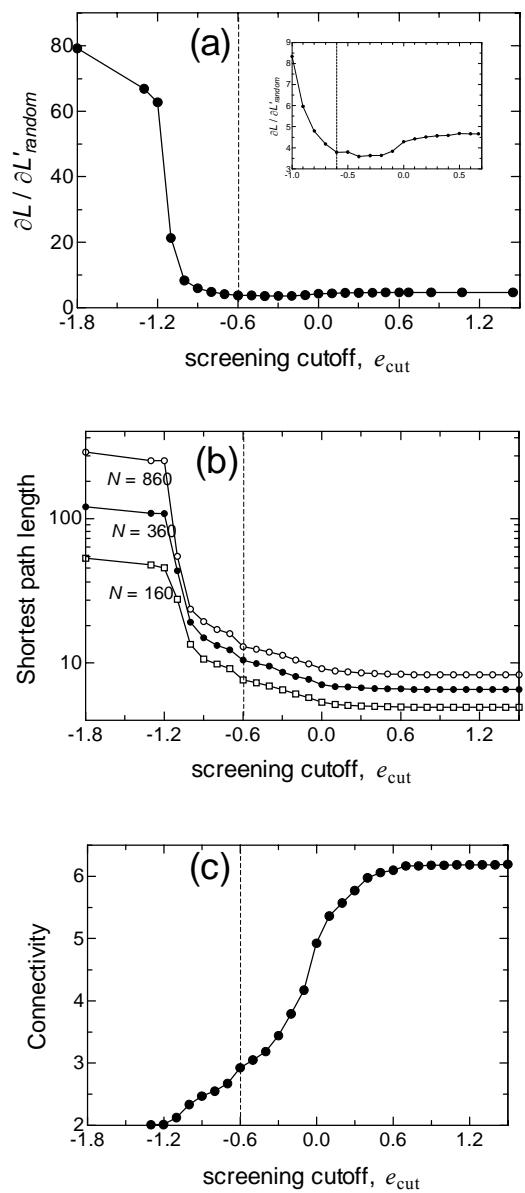

Figure 4

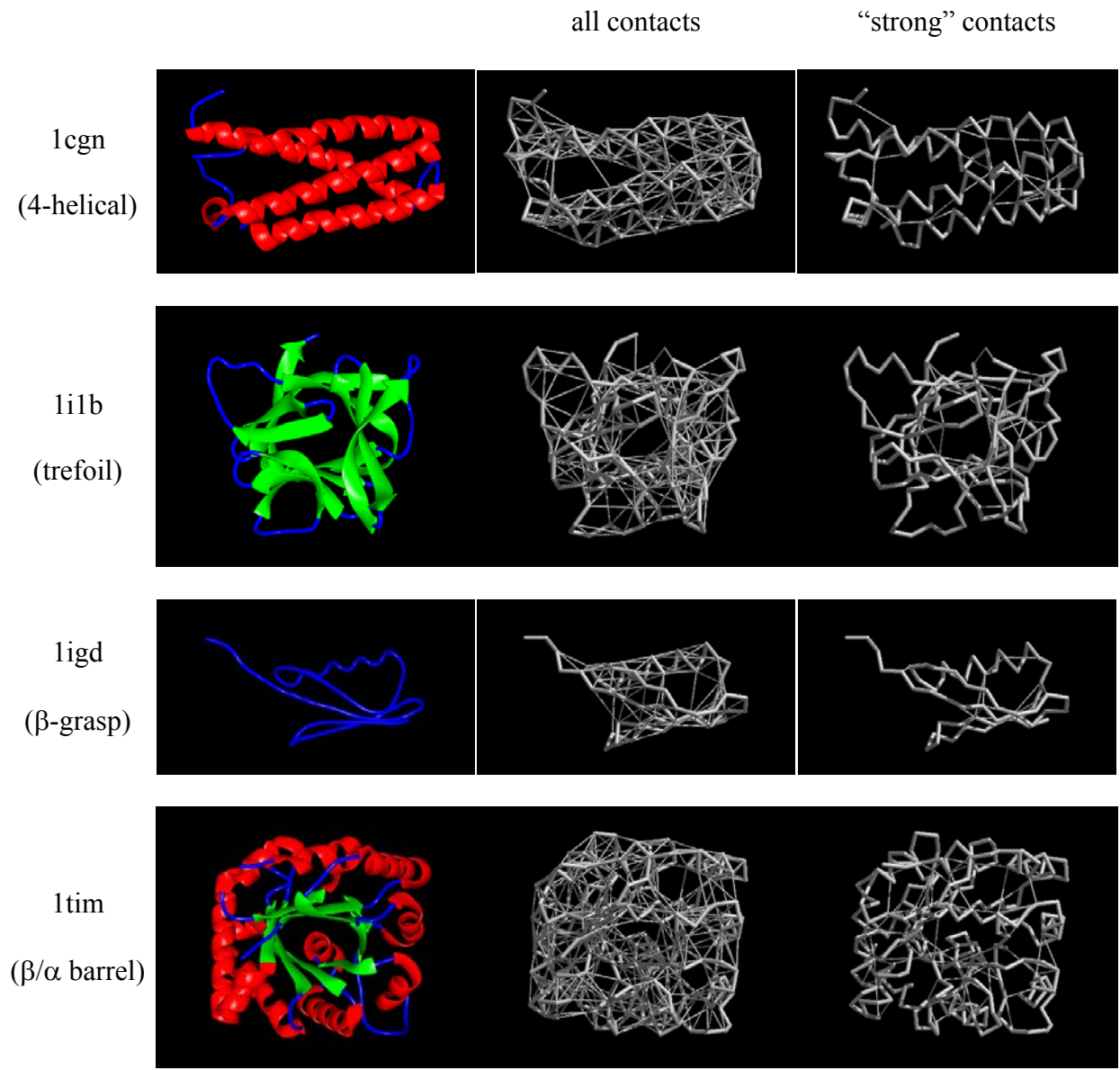

Figure 5

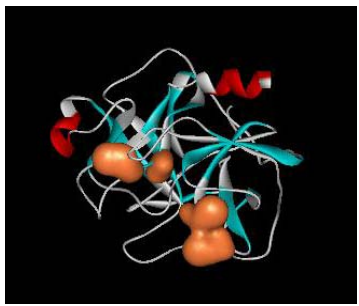

**RECEPTOR**

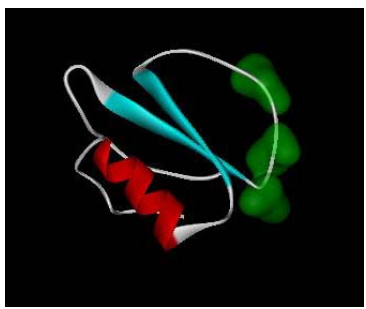

**LIGAND**

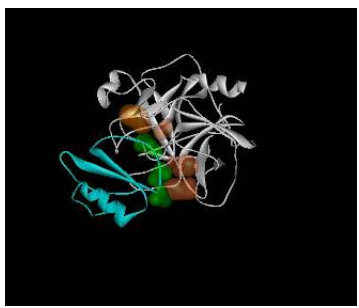

**COMPLEX**

Figure 6